\documentclass[epj,nopacs]{svjour}
\usepackage{graphicx}
\usepackage{mathtext}
\usepackage{amsmath}
\usepackage{amssymb}
\usepackage{color}

\onecolumn

\begin{document}
\title{Low dimensional dynamics of a sparse balanced synaptic network of quadratic integrate-and-fire neurons}
\titlerunning{Low dimensional dynamics of a sparse balanced network of QIF neurons}

\author{Maria~V.~Ageeva\inst{1,}\thanks{e-mail: ageeva\_mv00@mail.ru}
        \and Denis~S.~Goldobin\inst{1,2,
        }\thanks{e-mail: Denis.Goldobin@gmail.com}}
\authorrunning{M.~V.~Ageeva, D.~S.~Goldobin}

\institute{
Institute of Continuous Media Mechanics, UB RAS, Academician Korolev Street 1, 614013 Perm, Russia
\and Institute of Physics and Mathematics, Perm State University, Bukirev Street 15, 614990 Perm, Russia
}
\date{\today}

%
%
%
%
%
%
%

\abstract{Kinetics of a balanced network of neurons with a sparse grid of synaptic links is well representable by the stochastic dynamics of a generic neuron subject to an effective shot noise. The rate of delta-pulses of the noise is determined self-consistently from the probability density of the neuron states. Importantly, the most sophisticated (but robust) collective regimes of the network do not allow for the diffusion approximation, which is routinely adopted for a shot noise in mathematical neuroscience. These regimes can be expected to be biologically relevant. For the kinetics equations of the complete mean field theory of a homogeneous inhibitory network of quadratic integrate-and-fire neurons, we introduce circular cumulants of the genuine phase variable and derive a rigorous two cumulant reduction for both time-independent conditions and modulation of the excitatory current. The low dimensional model is examined with numerical simulations and found to be accurate for time-independent states and dynamic response to a periodic modulation deep into the parameter domain where the diffusion approximation is not applicable. The accuracy of a low dimensional model indicates and explains a low embedding dimensionality of the macroscopic collective dynamics of the network. The reduced model can be instrumental for theoretical studies of inhibitory-excitatory balanced neural networks.
}




\maketitle

\section{Introduction}
\label{sec1}
For balanced neural networks~\cite{Vreeswijk-Sompolinsky-1996,Kadmon-Sompolinsky-2015} with a sparse grid of synaptic links, the collective regimes were identified to be controlled by intrinsic fluctuations~\cite{Volo-Torcini-2018,Bi-Segneri-Volo-Torcini-2020,Volo-etal-2022,Goldobin-etal-2024,Goldobin-etal-2025}. These fluctuations are never negligible~\cite{Volo-Torcini-2018,Volo-etal-2022} and well representable by an effective Poissonian shot noise~\cite{Goldobin-etal-2024,Goldobin-etal-2025}. The diffusion approximation is conventionally adopted for shot-noise problems in mathematical neuroscience~\cite{Capocelli-Ricciardi-1971,Tuckwell-1988} and physics of condensed matter~\cite{Frisch-Lloyd-1960,Halperin-1965}. This approximation is mathematically accurate for a shot noise if the number of uncorrelated pulses received by a neuron per a macroscopic reference time ({\it or} spatial length) is large. The noise signal can be represented by two parts: the time-average value and the fluctuating part, which is white Gaussian noise in the case of the diffusion approximation.

The fluctuating part (white Gaussian noise) can be neglected in the thermodynamic limit for several paradigmatic problems. For these cases and for the problems with a Cauchy noise, the ``next-generation neural mass models''~\cite{Coombes-Byrne-2019,Luke-Barreto-So-2013,Laing-2014,Laing-2015,Pazo-Montbrio-2014,Montbrio-Pazo-Roxin-2015,Ratas-Pyragas-2016,Klinshov-Kirillov-2022,Pietras-Clusella-Montbrio-2025,Laing-Omelchenko-2023,Cestnik-Martens-2024,Pazo-Cestnik-2025,Toenjes-Pikovsky-2022,Pietras-etal-2023,Pyragas2-2023} were developed on the basis of the Ott--Antonsen theory~\cite{Ott-Antonsen-2008,Ott-Antonsen-2009} and allowed for a significant theoretical progress. Later on, on the basis of the circular and pseudo- cumulant approaches~\cite{Tyulkina-etal-2018,Goldobin-etal-2018,Goldobin-Dolmatova-2019,Goldobin-Dolmatova-2020,Goldobin-Volo-Torcini-2021}, upgraded versions of these neural mass models were developed to incorporate the white Gaussian noise~\cite{Volo-etal-2022,Goldobin-Volo-Torcini-2021,Ratas-Pyragas-2019,Zheng-Kotani-Jimbo-2021,Goldobin-2021,Pyragas-Pyragas-2024}.

For balanced networks with sparse grid of synaptic links, rich and nontrivial collective dynamics with important biological implications were recently reported~\cite{Goldobin-etal-2024,Goldobin-etal-2025} beyond the applicability limits of the diffusive approximation. Neither ``next-generation neural mass models'' nor their upgraded versions for Gaussian noise can be employed for this case. In this paper we derive low dimensional model reductions on the basis of circular cumulants specifically for a population of quadratic integrate-and-fire neurons (QIFs) and a shot noise.

%
%

We consider a dynamically balanced network of $N$ pulse-coupled QIFs with a sparse grid of inhibitory synaptic links. Membrane potentials $V_j$ of QIFs evolve according to the following equations~\cite{Ermentrout-Kopell-1986,Izhikevich-2007}:
\begin{align}
&\dot{V}_j=V_j^2+I-a\sum_{k=1}^{N}\sum_{n} \epsilon_{jk}\delta(t-t_k^{(n)})\;,
\label{eqGPi01}
\end{align}
where the adjacency matrix element $\epsilon_{jk}$ is $1$ if a synaptic link from the $k$-th neuron to the $j$-th one exists and $0$ otherwise; $K_j=\sum_k\epsilon_{jk}$ is the in- degree of the $j$-th neuron, and we consider a homogeneous population with identical $K$. $I=i_0\sqrt{K}$ represents an external DC current, $a=g_0/\sqrt{K}$: the synaptic coupling, $t_k^{(n)}$: the time of the $n$-th firing of the $k$-th neuron, and the last term: the inhibitory synaptic current. Here we explicitly indicate the scaling with $K$ required for the dynamical balance.
For a sufficiently sparse network with in- degree $K \ll N$ the spike train can be assumed to be uncorrelated and Poissonian. In this case the mean field dynamics of a generic QIF can be represented in the terms of the following Langevin equation:
\begin{align}
&\dot{V}=V^2+I-aS(t),
\label{eqGPi02}
\end{align}
where $S(t)$ is a Poissonian train of $\delta$ spikes with rate $R(t)=K\nu(t)$ and $\nu(t)$ is the population firing rate. For a homogeneous population within the mean-field framework the population dynamics can be described in terms of the membrane potential probability density function $P(V,t)$, whose time evolution is given by the continuity equation:
\begin{align}
&\frac{\partial P(V,t)}{\partial t}
=-\frac{\partial}{\partial V}\Big[\big(i_0\sqrt{K}+V^2\big)P(V,t)\Big]
+K\nu\Big[P(V+a,t)-P(V,t)\Big]\,,
\qquad
a=\frac{g_0}{\sqrt{K}}\,.
\label{eqSN201}
\end{align}
Below in the paper we deal with the dynamics of this equation.

The paper is organized as follows. In Section~\ref{sec2}, we present a detailed derivation of the complete mean field model---an infinite chain of equations for the dynamics of the Kuramoto--Daido order parameters of the genuine phase. In Section~\ref{sec3}, a rigorous two circular cumulant truncation of the infinite equation chain is derived. In Section~\ref{sec4}, we explicitly show that, for the shot noise, the relation between the firing rate, the mean membrane voltage and the probability density is still given by the same conventional Montbri\'o--Paz\'o--Roxin order parameter~\cite{Montbrio-Pazo-Roxin-2015}.
In Section~\ref{sec5}, we report final dimensionless models controlled by only two dimensionless parameters and validate the 2CC model reduction with the results of direct numerical simulation for time-independent regimes.
In Section~\ref{sec6}, we generalize mathematical models to the case of time-dependent modulation of parameters and present numerical results for a dynamic response for resonant and off-resonance modulation frequencies.
In Section~\ref{sec7}, we derive the diffusion approximation version of our mathematical models (complete system and 2CC reduction for a time-dependent modulation); in Section~\ref{sec71}, we suggest a theoretical estimate for the limits of applicability of the diffusion approximation.
In Section~\ref{sec:concl}, we finalize the paper with conclusion.

\section{Continuity equation for the probability density of genuine phase}
\label{sec2}
%
In the literature~\cite{Volo-etal-2022,Goldobin-etal-2024,Goldobin-etal-2025} it was reported and explained that for the network we consider any self-organized activity can arise only for the case of $i_0>0$. Hence, we can restrict our consideration to the case of $i_0>0$ and introduce genuine phase~\cite{genphase-2007,genphase-2008,Pikovsky-Rosenblum-Kurths-2003}, which is needed for a reliable detection of the synchronization level~\cite{Dolmatova-Goldobin-Pikovsky-2017,Goldobin-etal-2024,Goldobin-etal-2025}:
\[
\psi=2\arctan\frac{V}{\sqrt{I_0}}\,,\qquad
V=\sqrt{I_0}\tan\frac{\psi}{2}\,,
\]
where $I_0=i_0\sqrt{K}$.
The genuine phase representation (as in~\cite{Goldobin-etal-2024,Goldobin-etal-2025}) is more preferable for us than the usage of a conventional ``protophase'' $\theta=2\arctan{V}$ (as in, e.g.,~\cite{Volo-etal-2022,Ratas-Pyragas-2019}).
For the derivations in Sections~\ref{sec3}, \ref{sec6}, and~\ref{sec7}
and also for a didactic reason, in this Section we provide a detailed derivation of the mathematical model reported in~\cite{Goldobin-etal-2024,Goldobin-etal-2025}.

With
\[
\mathrm{d}V=\frac{\sqrt{I_0}}{2}\left(1+\tan^2\frac{\psi}{2}\right)\mathrm{d}\psi\,,
\qquad
P(V)\left|\mathrm{d}V\right|=w(\psi)\left|\mathrm{d}\psi\right|\,,
\qquad
P(V)=w(\psi)\frac{2\sqrt{I_0}}{I_0+V^2}\,,
\]
continuity equation~(\ref{eqSN201}) can be recast as
\begin{align}
&\frac{\partial w(\psi,t)}{\partial t}
=-\frac{\partial}{\partial\psi}\left[2\sqrt{I_0}w(\psi,t)\right]
+K\nu\left[\frac{I_0+V^2}{I_0+(V+a)^2}w(\psi_a,t)-w(\psi,t)\right]\,,
\label{eqGP01}
\end{align}
where
\[
V+a=\sqrt{I_0}\tan\frac{\psi_a}{2}\,,
\qquad
\tan\frac{\psi_a}{2}=\alpha+\tan\frac{\psi}{2}\,,
\qquad
\psi_a=2\arctan\left(\alpha+\tan\frac{\psi}{2}\right)\,,
\qquad \alpha\equiv\frac{a}{\sqrt{I_0}}\,.
\]
Eq.~(\ref{eqGP01}) can be finally written in terms of $\psi$ as
\begin{align}
&\frac{\partial w(\psi,t)}{\partial t}
=-\frac{\partial}{\partial\psi}\left[2\sqrt{I_0}w(\psi,t)\right]
+K\nu\left[\frac{w(\psi_a,t)}{1+\frac{\alpha^2}{2}+\alpha\sin\psi+\frac{\alpha^2}{2}\cos\psi} -w(\psi,t)\right]\,.
\label{eqGP02}
\end{align}

In Fourier space,
\[
w(\psi,t)=\frac{1}{2\pi}\left(1+\sum_{n=1}^{+\infty}z_ne^{-in\psi} +c.c.\right)\,,
\]
where ``$c.c.$'' stands for the complex conjugate,
or we can write
\[
w(\psi,t)=\frac{1}{2\pi}\sum\limits_{n=-\infty}^{+\infty}z_ne^{-in\psi},
\]
with $z_0=1$ and $z_{-n}=z_n^\ast$. Eq.~(\ref{eqGP02}) reads
\begin{align}
&\dot{z}_n=i2n\sqrt{I_0}z_n +K\nu\left[\sum_{m=-\infty}^{+\infty} I_{nm}z_m-z_n\right]\,,
\label{eqGP03}
\end{align}
where coefficients are given by the integrals
\begin{equation}
I_{nm}\equiv\frac{1}{2\pi}\int\limits_0^{2\pi} \frac{e^{in\psi}\left(e^{-i\psi_a}\right)^m\mathrm{d}\psi}{1+\frac{\alpha^2}{2}+\alpha\sin\psi+\frac{\alpha^2}{2}\cos\psi}\;.
\label{eqGP04}
\end{equation}

It is necessary to calculate
\begin{align}
e^{-i\psi_a}=&\cos\psi_a-i\sin\psi_a =\frac{1-i2\tan\frac{\psi_a}{2}-\tan^2\frac{\psi_a}{2}}{1+\tan^2\frac{\psi_a}{2}}
\nonumber\\
&=\frac{\left(1-i\tan\frac{\psi_a}{2}\right)^2}{\left(1+i\tan\frac{\psi_a}{2}\right)\left(1-i\tan\frac{\psi_a}{2}\right)}
 =\frac{1-i\tan\frac{\psi_a}{2}}{1+i\tan\frac{\psi_a}{2}}
=-\frac{\tan\frac{\psi}{2}+\alpha+i}{\tan\frac{\psi}{2}+\alpha-i}\;.
\nonumber
\end{align}
With (for $\alpha=0$)
\[
e^{-i\psi}=-\frac{\tan\frac{\psi}{2}+i}{\tan\frac{\psi}{2}-i}\;,
\quad\mbox{hence }
\tan\frac{\psi}{2}=i\frac{1-e^{i\psi}}{1+e^{i\psi}}\;,
\]
one can obtain
\begin{align}
e^{-i\psi_a}&=-\frac{\alpha +2i +\alpha e^{i\psi}}{\alpha +(\alpha-2i)e^{i\psi}}\;.
\label{eqGP05}
\end{align}

With~(\ref{eqGP05}) and substitution $e^{i\psi}=\zeta$, integral~(\ref{eqGP04}) can be rewritten as
\begin{align}
I_{nm}&=\frac{1}{2\pi}\int\limits_0^{2\pi}
 \frac{e^{in\psi}\left[-\frac{\alpha+2i+\alpha e^{i\psi}}{\alpha +(\alpha-2i)e^{i\psi}}\right]^m\mathrm{d}\psi}
 {1+\frac{\alpha^2}{2}-\frac{i\alpha}{2}(e^{i\psi}-e^{-i\psi})+\frac{\alpha^2}{4}(e^{i\psi}+e^{-i\psi})}
=\frac{1}{2\pi i}\oint\limits_{|\zeta|=1}
 \frac{\zeta^{n-1} \left[-\frac{\alpha+2i+\alpha\zeta}{\alpha+(\alpha-2i)\zeta}\right]^m\mathrm{d}\zeta}
 {1+\frac{\alpha^2}{2} -\frac{i\alpha}{2}(\zeta-\frac{1}{\zeta}) +\frac{\alpha^2}{4}(\zeta+\frac{1}{\zeta})}
\nonumber\\
&=\frac{1}{2\pi i}\oint\limits_{|\zeta|=1}
 \frac{\frac{\zeta^n}{(\alpha-2i)^m}\left[-\frac{\alpha+2i+\alpha\zeta}
 {\zeta+\frac{\alpha}{\alpha-2i}}\right]^m\mathrm{d}\zeta}{(\frac{\alpha^2}{4}-\frac{i\alpha}{2})\zeta^2+(1+\frac{\alpha^2}{2})\zeta +\frac{\alpha^2}{4}+\frac{i\alpha}{2}}
=\frac{1}{2\pi i}\oint\limits_{|\zeta|=1}
 \frac{ \frac{4\zeta^n}{\alpha(\alpha-2i)^{m+1}}\left[-\frac{\alpha+2i+\alpha\zeta}
 {\zeta+\frac{\alpha}{\alpha-2i}}\right]^m\mathrm{d}\zeta}
 {(\zeta+\frac{\alpha}{\alpha-2i})(\zeta+\frac{\alpha+2i}{\alpha})}
\nonumber\\
&=\frac{1}{2\pi i}\oint\limits_{|\zeta|=1}
 \frac{4(-\alpha)^m\zeta^n(\zeta+\frac{\alpha+2i}{\alpha})^{m-1}}
 {\alpha(\alpha-2i)^{m+1}(\zeta+\frac{\alpha}{\alpha-2i})^{m+1}}\mathrm{d}\zeta\;.
\label{eqGP06}
\end{align}
It is enough to consider $n\ge1$, as there is no dynamics of $z_0=1$, and for negative orders $z_{-n}=z_n^\ast$.
For $n\ge1$, the integrand numerator can possess poles at
\[
\zeta_2=-\frac{\alpha+2i}{\alpha}\,,
\]
which are always beyond the integration contour $|\zeta|=1$ as $|(\alpha+2i)/\alpha|^2=1+4/\alpha^2>1$ and, therefore, do not contribute to the integral $I_{nm}$. The integrand has poles at
\[
\zeta_1=-\frac{\alpha}{\alpha-2i}
\]
for $m>0$\,. These poles are always within the integration contour as $|\alpha/(\alpha-2i)|^2=\alpha^2/(\alpha^2+4)<1$.

Now we can employ the Residue theorem and calculate
\begin{align}
&\frac{1}{2\pi i}\oint\limits_{|\zeta|=1}
 \frac{\zeta^n(\zeta+\frac{\alpha+2i}{\alpha})^{m-1}}
 {(\zeta+\frac{\alpha}{\alpha-2i})^{m+1}}\mathrm{d}\zeta
=\left\{
\begin{array}{cr}
\frac{1}{m!}\left.\frac{\mathrm{d}^m}{\mathrm{d}\zeta^m} \left(\zeta^n\left(\zeta-\zeta_2\right)^{m-1}\right)\right|_{\zeta=\zeta_1}\;,
& m\ge0\;;\\
0\;,
& m\le-1
\end{array}
\right.
\nonumber\\
&\qquad
=\left\{
\begin{array}{cr}
\sum\limits_{j=0}^{\min(n,m)-1}\Big({m-1 \atop j}\Big)\Big({n+m-1-j \atop m}\Big)
\zeta_1^{n-1-j}\left(-\zeta_2\right)^{j}\;,
&m\ge 1\;;\\[5pt]
\frac{\zeta_1^n}{\zeta_1-\zeta_2}\;,
&m=0\;;\\[5pt]
0\;,
&m\le-1
\end{array}
\right.
\nonumber
\end{align}
where $\min(n,m)$ stands for the minimal of two values, the binomial coefficients $\left(n\atop m\right)=\frac{n!}{m!(n-m)!}$;
for obtaining the line for $m\ge1$ one should separately consider two cases: $1\le n\le m$ and $n>m$. For convenience, we introduce
\[
I_{nm}\equiv\frac{4}{\alpha(\alpha-2i)}\mathcal{I}_{nm}\,,
\]
\begin{align}
\mathcal{I}_{nm}=\left\{
\begin{array}{cr}
\sum\limits_{j=0}^{\min(n,m)-1}
\frac{(n+m-1-j)!\,\zeta_1^{m+n-1-j}\left(-\zeta_2\right)^{j}}{m\, j!\,(m-1-j)!\,(n-1-j)!},
&m\ge 1\,;\\[5pt]
\frac{\zeta_1^{n}}{\zeta_1-\zeta_2}\,,
&m=0\,;\\[5pt]
0\,,
&m\le-1\,.
\end{array}
\right.
\nonumber
\end{align}

For $n\ge1$, one finds $\mathcal{I}_{nm}=0$ for $m<0$; therefore, we deal with the matrix $\mathcal{I}_{nm}$ (or $I_{nm}$) only for $n=1,2,3,...$ and $m=0,1,2,3,...$\,:
\begin{align}
&(\mathcal{I}_{nm})=
\left(
\begin{array}{llllll}
\frac{\zeta_1}{\zeta_1-\zeta_2} & \zeta_1 & \zeta_1^2 & \zeta_1^3 & \zeta_1^4
&\dots
\\[5pt]
\frac{\zeta_1^2}{\zeta_1-\zeta_2} & 2\zeta_1^2 & 3\zeta_1^3-\zeta_1^2\zeta_2 & 4\zeta_1^4-2\zeta_1^3\zeta_2 & 5\zeta_1^5-3\zeta_1^4\zeta_2
&\dots
\\[5pt]
\frac{\zeta_1^3}{\zeta_1-\zeta_2} & 3\zeta_1^3 & 6\zeta_1^4-3\zeta_1^3\zeta_2 & 10\zeta_1^5-8\zeta_1^4\zeta_2+\zeta_1^3\zeta_2^2
& 15\zeta_1^6-15\zeta_1^5\zeta_2+3\zeta_1^4\zeta_2^2
&\dots
\\[5pt]
\frac{\zeta_1^4}{\zeta_1-\zeta_2} \; & 4\zeta_1^4 \; & 10\zeta_1^5-6\zeta_1^4\zeta_2 \; & 20\zeta_1^6-20\zeta_1^5\zeta_2+4\zeta_1^4\zeta_2^2 \;
& 35\zeta_1^7-45\zeta_1^6\zeta_2+15\zeta_1^5\zeta_2^2-\zeta_1^4\zeta_2^3 \;
&\dots
\\[5pt]
&\dots
\end{array}
\right).
\label{eqGP07}
\end{align}

\section{Two circular cumulant reduction for genuine phase}
\label{sec3}
Let us now switch from the circular moments $\{z_n\}$ to ``circular cumulants'' (CC)~\cite{Tyulkina-etal-2018,Goldobin-etal-2018,Goldobin-Dolmatova-2019}
\[
\varkappa_n=\frac{z_n}{(n-1)!} -\sum_{m=1}^{n-1}\frac{\varkappa_mz_{n-m}}{(n-m)!};
\]
in particular, two first cumulants are
\[
\varkappa_1=z_1\,,\qquad
\varkappa_2=z_2-z_1^2\,,
\]
where the first CC is identical to the Kuramoto order parameter and the second CC measures the deviation of the second circular moment $z_2$ from the Ott--Antonsen manifold $z_n=(z_1)^n$~\cite{Ott-Antonsen-2008,Ott-Antonsen-2009}.
With a two circular cumulant reduction, the characteristic function $1+z_1k+z_2\frac{k^2}{2!}+z_3\frac{k^3}{3!}+z_4\frac{k^4}{4!}+\dots\approx e^{\varkappa_1k+\varkappa_2\frac{k^2}{2}} \approx (1+\varkappa_2\frac{k^2}{2})e^{\varkappa_1k}$; therefore, $z_m\approx\varkappa_1^m+\frac{m(m-1)}{2}\varkappa_2\varkappa_1^{m-2}$~\cite{Tyulkina-etal-2018,Goldobin-etal-2018,Goldobin-Dolmatova-2019}.
Hence,
\begin{align}
\sum_{m=0}^{\infty}\mathcal{I}_{1m}z_m &= \frac{\zeta_2}{\zeta_1-\zeta_2} +\sum_{m=0}^{\infty}\zeta_1^m\left(\varkappa_1^m+\frac{m(m-1)}{2}\varkappa_2\varkappa_1^{m-2}\right)
=\frac{\zeta_2}{\zeta_1-\zeta_2} +\left(1+\frac{\varkappa_2}{2}\frac{\partial^2}{\partial\varkappa_1^2}\right) \sum_{m=0}^{\infty}\zeta_1^m\varkappa_1^m
\nonumber\\
&
=\frac{\zeta_2}{\zeta_1-\zeta_2} +\left(1+\frac{\varkappa_2}{2}\frac{\partial^2}{\partial\varkappa_1^2}\right) \frac{1}{1-\zeta_1\varkappa_1}
=\frac{\zeta_2}{\zeta_1-\zeta_2} +\frac{1}{1-\zeta_1\varkappa_1} +\frac{\varkappa_2\zeta_1^2}{(1-\zeta_1\varkappa_1)^3}\,.
\nonumber
\end{align}
Since $\dot\varkappa_2=\dot{z}_2-2z_1\dot{z}_1$\,, we need, for $m\ge1$,
\begin{align}
\mathcal{I}_{2m}z_m-2z_1\mathcal{I}_{1m}z_m &=\left[(m+1)\zeta_1^{m+1}-(m-1)\zeta_1^m\zeta_2\right]z_m-2z_1\zeta_1^mz_m
\nonumber\\
&=\left(1+\frac{\varkappa_2}{2}\frac{\partial^2}{\partial\varkappa_1^2}\right)
\left[(m+1)\zeta_1^{m+1}-(m-1)\zeta_1^m\zeta_2\right]\varkappa_1^m
-2\varkappa_1\left(1+\frac{\varkappa_2}{2}\frac{\partial^2}{\partial\varkappa_1^2}\right)\zeta_1^m\varkappa_1^m\,;
\nonumber
\end{align}
and, for $m=0$,
\[
\mathcal{I}_{20}z_0-2z_1\mathcal{I}_{10}z_0=\frac{\zeta_1^2}{\zeta_1-\zeta_2} -\frac{2\zeta_1\varkappa_1}{\zeta_1-\zeta_2}\,.
\]
Hence,
\begin{align}
&\sum_{m=0}^{\infty}\left(\mathcal{I}_{2m}z_m-2z_1\mathcal{I}_{1m}z_m\right)
=\frac{\zeta_1^2-2\zeta_1\varkappa_1}{\zeta_1-\zeta_2}
\nonumber\\
&\qquad\qquad\qquad
+\sum_{m=1}^{\infty}\left\{\left(1+\frac{\varkappa_2}{2}\frac{\partial^2}{\partial\varkappa_1^2}\right)
\left[(m+1)\zeta_1^{m+1}-(m-1)\zeta_1^m\zeta_2\right]\varkappa_1^m
-2\varkappa_1\left(1+\frac{\varkappa_2}{2}\frac{\partial^2}{\partial\varkappa_1^2}\right)\zeta_1^m\varkappa_1^m\right\}
\nonumber\\
&\quad
=\frac{\zeta_1^2-2\zeta_1\varkappa_1}{\zeta_1-\zeta_2} +\left(1+\frac{\varkappa_2}{2}\frac{\partial^2}{\partial\varkappa_1^2}\right)
\left(\frac{\partial}{\partial\varkappa_1}\frac{\zeta_1^2\varkappa_1^2}{1-\zeta_1\varkappa_1}
-\zeta_1\zeta_2\varkappa_1^2\frac{\partial}{\partial\varkappa_1}\frac{1}{1-\zeta_1\varkappa_1}\right)
-2\varkappa_1\left(1+\frac{\varkappa_2}{2}\frac{\partial^2}{\partial\varkappa_1^2}\right) \frac{\zeta_1\varkappa_1}{1-\zeta_1\varkappa_1}
\nonumber\\
&\quad
=\frac{\zeta_1^2-2\zeta_1\varkappa_1}{\zeta_1-\zeta_2} +\left(1+\frac{\varkappa_2}{2}\frac{\partial^2}{\partial\varkappa_1^2}\right)
\left(\frac{2\zeta_1^2\varkappa_1}{1-\zeta_1\varkappa_1}
+\frac{\zeta_1^2\varkappa_1^2(\zeta_1-\zeta_2)}{(1-\zeta_1\varkappa_1)^2}\right)
-2\varkappa_1\left(1+\frac{\varkappa_2}{2}\frac{\partial^2}{\partial\varkappa_1^2}\right) \frac{\zeta_1\varkappa_1}{1-\zeta_1\varkappa_1}
\nonumber\\
&\quad
=\frac{\zeta_1^2-2\zeta_1\varkappa_1}{\zeta_1-\zeta_2} +\frac{2\zeta_1\varkappa_1(\zeta_1-\varkappa_1)}{1-\zeta_1\varkappa_1}
+\frac{\zeta_1^2\varkappa_1^2(\zeta_1-\zeta_2)}{(1-\zeta_1\varkappa_1)^2}
+\varkappa_2\left(\frac{2\zeta_1^2(\zeta_1-\varkappa_1)}{(1-\zeta_1\varkappa_1)^3}
+\zeta_1^2(\zeta_1-\zeta_2)\frac{1+2\zeta_1\varkappa_1}{(1-\zeta_1\varkappa_1)^4}
\right).
\nonumber
\end{align}
Therefore, two first equations of the infinite chain of CC equations corresponding to (\ref{eqGP03}) can be written down:
\begin{align}
&\dot\varkappa_1=i2\sqrt{I_0}\varkappa_1 +K\nu\left[\frac{4}{\alpha(\alpha-2i)}\left\{\frac{\zeta_2}{\zeta_1-\zeta_2} +\frac{1}{1-\zeta_1\varkappa_1} +\frac{\varkappa_2\zeta_1^2}{(1-\zeta_1\varkappa_1)^3}\right\} -\varkappa_1\right],
\label{eqGP08}
\\
&\dot\varkappa_2=i4\sqrt{I_0}\varkappa_2 +K\nu\left[\frac{4}{\alpha(\alpha-2i)}\left\{
\frac{\zeta_1(\zeta_1-2\varkappa_1)}{\zeta_1-\zeta_2} +\frac{2\zeta_1\varkappa_1(\zeta_1-\varkappa_1)}{1-\zeta_1\varkappa_1}
+\frac{\zeta_1^2\varkappa_1^2(\zeta_1-\zeta_2)}{(1-\zeta_1\varkappa_1)^2}
\right.\right.
\nonumber\\
&\qquad\qquad\qquad\qquad\qquad\quad\left.\left.
{}
+\varkappa_2\left(\frac{2\zeta_1^2(\zeta_1-\varkappa_1)}{(1-\zeta_1\varkappa_1)^3}
+\zeta_1^2(\zeta_1-\zeta_2)\frac{1+2\zeta_1\varkappa_1}{(1-\zeta_1\varkappa_1)^4}
\right)\right\}
+\varkappa_1^2-\varkappa_2\right].
\label{eqGP09}
\end{align}
One can calculate $\zeta_1-\zeta_2=4/[\alpha(\alpha-2i)]$,  $\zeta_2=-2-1/\zeta_1$ and finally write:
\begin{align}
&\dot\varkappa_1=i2K^\frac14\sqrt{i_0}\varkappa_1
 +K\nu\left[
 \frac{\zeta_1(1+\varkappa_1)^2}{1-\zeta_1\varkappa_1} +\frac{\zeta_1(1+\zeta_1)^2}{(1-\zeta_1\varkappa_1)^3}\varkappa_2\right],
\label{eqGP10}
\\
&\dot\varkappa_2=i4K^\frac14\sqrt{i_0}\varkappa_2 +K\nu\left[
\frac{\zeta_1^2(1+\varkappa_1)^4}{(1-\zeta_1\varkappa_1)^2}
-\varkappa_2\left(1 -\frac{2(1+\zeta_1)^2}{(1-\zeta_1\varkappa_1)^2}
+\frac{4(1+\zeta_1)^3}{(1-\zeta_1\varkappa_1)^3} -\frac{3(1+\zeta_1)^4}{(1-\zeta_1\varkappa_1)^4}\right)
\right].
\label{eqGP11}
\end{align}

\section{Relation between firing rate, mean membrane potential and probability density of the genuine phase}
\label{sec4}
In Eq.~(\ref{eqGP01}) the deterministic part of the probability flux $q_\psi$ is $2\sqrt{I_0}\,w(\psi,t)$ and the noise-driven part vanishes for $\psi\to\pm\pi$, since $\lim_{\psi\to\pm\pi}\frac{\psi_a-\psi}{a}=0$. Therefore,
\begin{align}
\nu&=q_\psi(\pi)=2\sqrt{I_0}\,w(\pi)
=\frac{\sqrt{I_0}}{\pi}\mathrm{Re}(1-2z_1+2z_2-2z_3+2z_4-\dots)
\label{eqGP12zn}
\\
&=\frac{K^\frac14\sqrt{i_0}}{\pi}\mathrm{Re}\left(\frac{1-\varkappa_1} {1+\varkappa_1}+\frac{2\varkappa_2}{(1+\varkappa_1)^3}+\dots\right).
\label{eqGP12}
\end{align}
Further, the population mean voltage
\begin{align}
v&=\mathrm{P.V.}\int\limits_{-\infty}^{+\infty} VP(V,t)\mathrm{d}V
=\int\limits_{-\pi+0}^{\pi-0}\sqrt{I_0}\tan\frac{\psi}{2}w(\psi,t)\mathrm{d}\psi
=-\sqrt{I_0}\mathrm{Im}(1-2z_1+2z_2-2z_3+2z_4-\dots)
\nonumber\\
&=-K^\frac14\sqrt{i_0}\mathrm{Im}\left(\frac{1-\varkappa_1} {1+\varkappa_1}+\frac{2\varkappa_2}{(1+\varkappa_1)^3}+\dots\right)
\label{eqGPv}
\end{align}
(the second line is straightforwardly calculated from the integral as in Ref.~\cite{Montbrio-Pazo-Roxin-2015}).
Altogether, one can write
\begin{align}
\label{eqGPW}
&\pi\nu-iv=\sqrt{I_0}W_\psi
\equiv\sqrt{I_0}(1-2z_1+2z_2-2z_3+2z_4-\dots)\;,
\end{align}
where we use subscript to explicitly indicate that $W_\psi$ is taken for the genuine phase $\psi$.

\section{Circular cumulant equations with rescaled time}
\label{sec5}
Notice that firing rate~(\ref{eqGP12}) and the first terms in (\ref{eqGP10})--(\ref{eqGP11}) contain the same prefactor $K^{1/4}\sqrt{i_0}$\,. In terms of
\begin{align}
\widetilde{\nu}&=\frac{\nu}{K^{1/4}\sqrt{i_0}}
=\frac{1}{\pi}\mathrm{Re}(1-2z_1+2z_2-2z_3+2z_4-\dots)
\nonumber\\
&\approx\frac{1}{\pi}\mathrm{Re}\left(\frac{1-\varkappa_1} {1+\varkappa_1}+\frac{2\varkappa_2}{(1+\varkappa_1)^3}+\dots\right)\,,
\label{eqGPnutilda}
\end{align}
with rescaled time $\tau=K^{1/4}\sqrt{i_0}\,t$, one can rewrite (\ref{eqGP10})--(\ref{eqGP11}) as
\begin{align}
&\frac{\mathrm{d}\varkappa_1}{\mathrm{d}\tau}=i2\varkappa_1 +K\widetilde{\nu}\left[
 \frac{\zeta_1(1+\varkappa_1)^2}{1-\zeta_1\varkappa_1} +\frac{\zeta_1(1+\zeta_1)^2}{(1-\zeta_1\varkappa_1)^3}\varkappa_2\right],
\label{eqGP13}
\\
&\frac{\mathrm{d}\varkappa_2}{\mathrm{d}\tau}=i4\varkappa_2 +K\widetilde{\nu}\left[
\frac{\zeta_1^2(1+\varkappa_1)^4}{(1-\zeta_1\varkappa_1)^2}
-\varkappa_2\left(1 -\frac{2(1+\zeta_1)^2}{(1-\zeta_1\varkappa_1)^2}
+\frac{4(1+\zeta_1)^3}{(1-\zeta_1\varkappa_1)^3} -\frac{3(1+\zeta_1)^4}{(1-\zeta_1\varkappa_1)^4}\right)
\right].
\label{eqGP14}
\end{align}
Noteworthy, the dynamical system (\ref{eqGPnutilda})--(\ref{eqGP14}) is not independent of $i_0$ as $\alpha=g_0/(K^{3/4}\sqrt{i_0})$. Thus, the dynamics of the 2CC reduction model, as well as the dynamics of the original model (\ref{eqGP03}) and (\ref{eqGP12zn})~\cite{Goldobin-etal-2024,Goldobin-etal-2025}, is controlled by only two dimensionless parameters: $K$ and $i_0/g_0^2$ (or $\alpha$). Thus, the macroscopic dynamics of the network can be comprehensively presented on the parameter plane $(K,i_0/g_0^2)$.

\begin{figure}[!t]
\center{
\includegraphics[width=0.32\textwidth]%
 {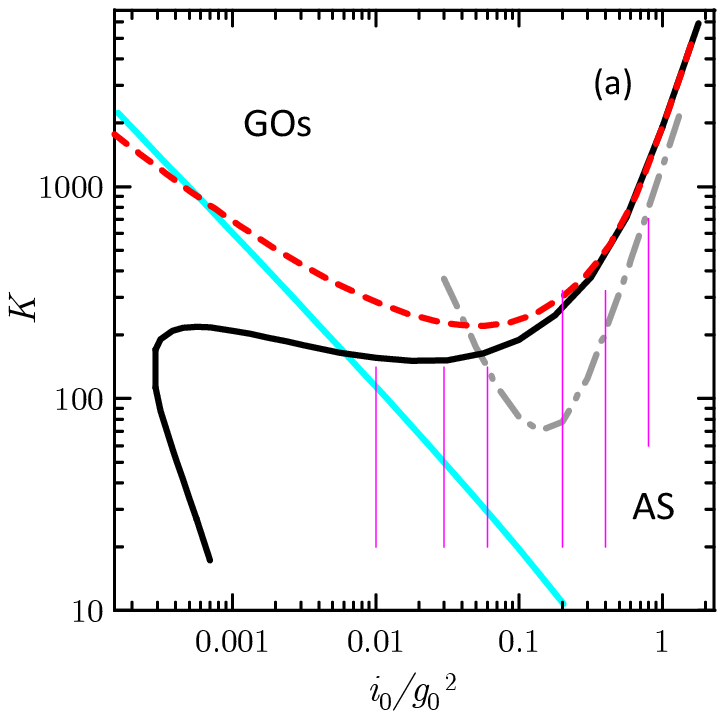}
 \hspace{1.5mm}
\includegraphics[width=0.32\textwidth]%
 {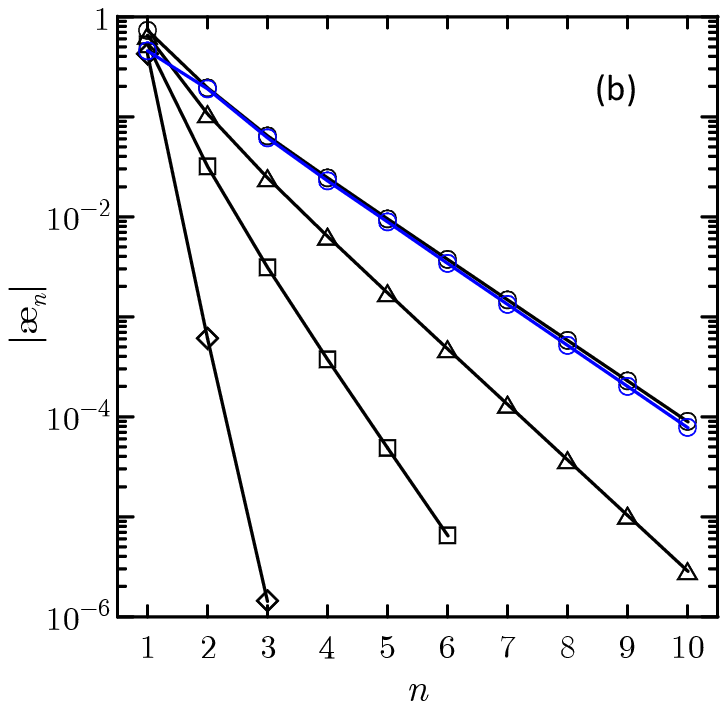}
 \hspace{1.5mm}
\includegraphics[width=0.3125\textwidth]%
 {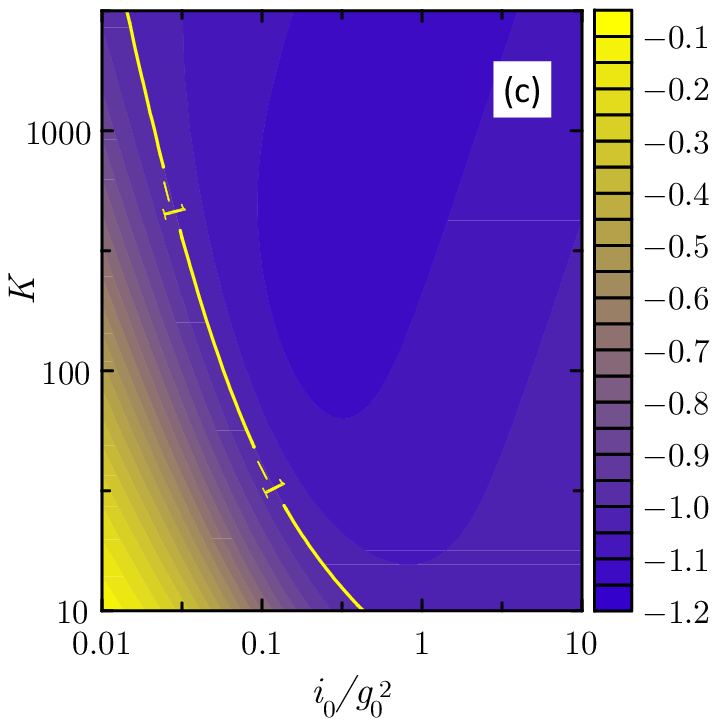}
}
\caption{(a):~The diagram of the macroscopic regimes (GOs: global oscillations, AS: asynchronous dynamics). The boundary between the GO and AS regimes is the Hopf bifurcation curve plotted with the black solid line for the original complete mean field system (\ref{eqGP03}) and (\ref{eqGP12zn}) and with the red dashed line for the diffusion approximation (Section~\ref{sec7}). The cyan line: $K_\mathrm{D}(i_0/g_0^2)$, the diffusion approximation is accurate for in- degree $K$ 1--2 orders of magnitude larger than $K_\mathrm{D}$ (Section~\ref{sec71}).
The gray dash-dotted line: the Hopf bifurcation curve for model reduction~(\ref{eqGPnutilda})--(\ref{eqGP14}).
(b):~Circular cumulants for the ``exact'' time-independent solution of~(\ref{eqGP03}) and (\ref{eqGP12zn}) at the Hopf bifurcation line for $i_0/g_0^2=1.78$ (diamonds), $0.178$ (squares), $0.0316$ (triangles), $0.0007$ (black/blue circles for the upper/lower branch).
(c): The Lyapunov exponent $\lambda$ of the time-independent solution within the Ott--Antonsen Ansatz [given by Eqs.~(\ref{eqGPnutilda})--(\ref{eqGP13}) with $\varkappa_2=0$] is presented with the shadowgraph of $2\pi\sqrt{i_0}K^{1/4}\lambda/g_0^2$.
}
  \label{fig1}
\end{figure}

In Figure~\ref{fig1}, the Hopf bifurcation curve of the time-independent state separates the macroscopic regime of synchronous oscillatory dynamics (``GO'': global oscillations), where firing rate oscillates in time, and the asynchronous regime (AS), where QIFs oscillate incoherently and the firing rate of the population is constant (in the thermodynamic limit of an infinite population). These regimes and their biological interpretations are thoroughly studied in~\cite{Goldobin-etal-2024,Goldobin-etal-2025}.
In panel~(b) of Figure~\ref{fig1}, one can see that CCs $\{\varkappa_n\}$ for ``exact'' solutions form well pronounced hierarchies of smallness.
Here and hereafter, the ``exact'' solutions are obtained by the direct numerical simulation of equation chain~(\ref{eqGP03}) and (\ref{eqGP12zn}) with $M=64$ modes $\{z_n\}$ by means of the exponential time differencing method~\cite{Cox-Matthews-2002,Hochbruck-Ostermann-2010,Permyakova-Goldobin-2025}. 
High numerical accuracy of the ``exact'' solution is thoroughly validated in~\cite{Goldobin-etal-2025}.
For very small values of $i_0$ (circles: $i_0/g_0^2=0.0007$), the smallness of $\varkappa_3$ can be insufficient and the 2CC reduction produces noticeable inaccuracy. For $i_0/g_0^2\sim0.03$ and larger currents, one can expect a decent accuracy of the 2CC reduction.

\begin{figure}[!t]
\center{
\includegraphics[width=0.48\textwidth]%
 {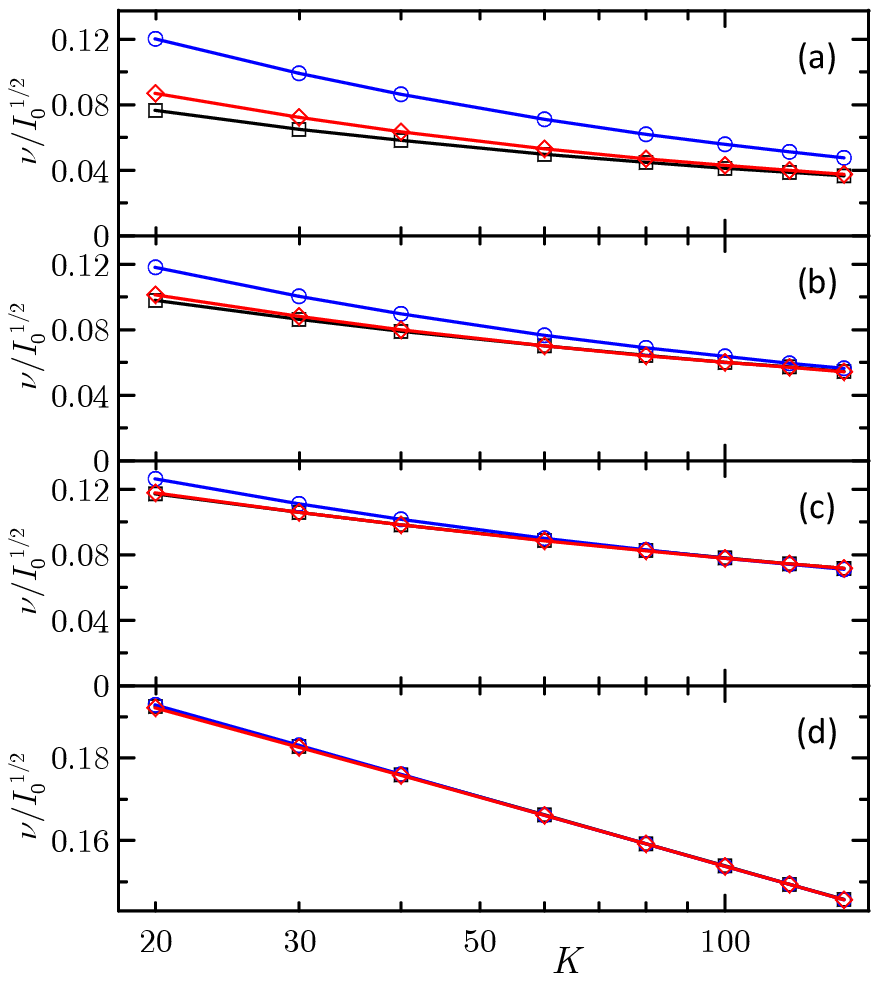}
}
\caption{The firing rate for time-independent network states is plotted for $i_0/g_0^2=0.01$ (a), $0.03$~(b), $0.06$~(c), and $0.4$~(d). Black squares: the ``exact'' solution of the infinite equation chain~(\ref{eqGP03}) with firing rate~(\ref{eqGP12zn}); red diamonds: 2CC reduction~(\ref{eqGPnutilda})--(\ref{eqGP14}); blue circles: Ott--Antonsen Ansatz given by Eqs.~(\ref{eqGPnutilda})--(\ref{eqGP13}) with $\varkappa_2=0$.
The values of $i_0$ and $K$ for these plots are shown with magenta lines in Figure~\ref{fig1}a.
}
  \label{fig2}
\end{figure}

In Figure~\ref{fig2}, one can see that 2CC reduction~(\ref{eqGPnutilda})--(\ref{eqGP14}) provides reasonable accuracy for the time-independent states (AS) for as low excitatory current as $i_0=0.01g_0^2$, and the accuracy rapidly becomes better as $i_0$ increases.
Here, for the sake of completeness, we also report the results for Ott--Antonsen (OA) Ansatz~\cite{Ott-Antonsen-2008,Ott-Antonsen-2009}, which is given by Eqs.~(\ref{eqGPnutilda})--(\ref{eqGP13}) with $\varkappa_2$ set to zero. One can see that the OA model reduction becomes accurate only for much higher values of the excitatory current $i_0$. Moreover, it is known to be fundamentally unable to reproduce the noise-induced oscillations in neural circuits within the diffusion approximation~\cite{Volo-Torcini-2018,Bi-Segneri-Volo-Torcini-2020,Volo-etal-2022,Goldobin-Volo-Torcini-2021,Ratas-Pyragas-2019}. For the OA reduction of the shot-noise model, the Lyapunov exponent $\lambda$ is also always negative as shown in Figure~\ref{fig1}c, where we explicitly indicate the asymptotic law which can be derived for large $i_0/g_0^2$ or large $K$: $\lambda_\infty\approx-g_0^2/(2\pi\sqrt{i_0}K^{1/4})$.
The applicability and performance of the 2CC model reduction for the case of time-dependent synaptic activity are examined in the next Section.

\section{Time-dependent external excitatory current $I(t)$}
\label{sec6}
One can introduce modulation of the external current $I$ as follows: $I(t)=I_0[1+\eta(t)]$; the genuine phase is defined with constant $I_0$ and modulation is given with $\eta(t)$. The continuity equation~(\ref{eqSN201}) with $I_0[1+\eta(t)]$ gives the evolution equation for $P(V,t)$. Substituting $w(\psi,t)=P(V,t)(I_0+V^2)/(2\sqrt{I_0})$, one finds in place of Eq.~(\ref{eqGP02}):
\begin{align}
\frac{\partial w(\psi,t)}{\partial t}
=-\frac{\partial}{\partial\psi}\left\{\sqrt{I_0}\big[2 +\eta(t)(1+\cos\psi)\big]w(\psi,t)\right\}
+K\nu(t) \left[\frac{w(\psi_a,t)}{1+\frac{\alpha^2}{2}+\alpha\sin\psi+\frac{\alpha^2}{2}\cos\psi} -w(\psi,t)\right]\,.
\label{eqGP02eta}
\end{align}
In Fourier space, the latter equation yields an extension of (\ref{eqGP03}):
\begin{align}
\dot{z}_n=
in\sqrt{I_0}\left\{\big[2+\eta(t)\big]z_n
 +\eta(t)\frac{z_{n-1}+z_{n+1}}{2}\right\}
 +K\nu(t)\left[\sum_{m=0}^{+\infty}I_{nm}\,z_m-z_n\right]\,;
\label{eqGP03eta}
\end{align}
The 2CC reduction for Eq.~(\ref{eqGP03eta}) requires the $\eta(t)$-terms to be incorporated into Eqs.~(\ref{eqGP13})--(\ref{eqGP14}):
\begin{align}
&\frac{\mathrm{d}\varkappa_1}{\mathrm{d}\tau}=2i\varkappa_1 +i\eta(\tau)\frac{(1+\varkappa_1)^2+\varkappa_2}{2} +K\widetilde{\nu}\left[
\frac{\zeta_1(1+\varkappa_1)^2}{1-\zeta_1\varkappa_1} +\frac{\zeta_1(1+\zeta_1)^2}{(1-\zeta_1\varkappa_1)^3}\varkappa_2\right],
\label{eqGPeta1}
\\
&\frac{\mathrm{d}\varkappa_2}{\mathrm{d}\tau}=4i\varkappa_2 +2i\eta(\tau)(1+\varkappa_1)\varkappa_2 +K\widetilde{\nu}\left[
\frac{\zeta_1^2(1+\varkappa_1)^4}{(1-\zeta_1\varkappa_1)^2}
-\varkappa_2\left(1 -\frac{2(1+\zeta_1)^2}{(1-\zeta_1\varkappa_1)^2}
+\frac{4(1+\zeta_1)^3}{(1-\zeta_1\varkappa_1)^3} -\frac{3(1+\zeta_1)^4}{(1-\zeta_1\varkappa_1)^4}\right)
\right].
\label{eqGPeta2}
\end{align}
The relation between the firing rate and $\{z_n\}$ remains unchanged.

\begin{figure}[!t]
\center{
\includegraphics[width=0.32\textwidth]%
 {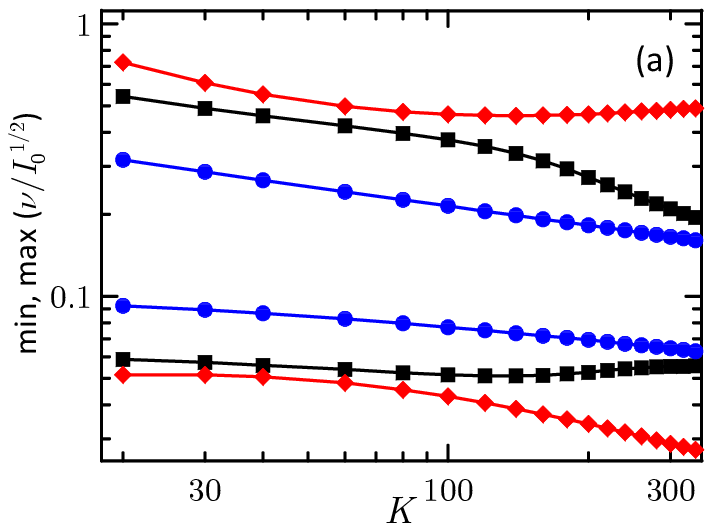}
\;
\includegraphics[width=0.32\textwidth]%
 {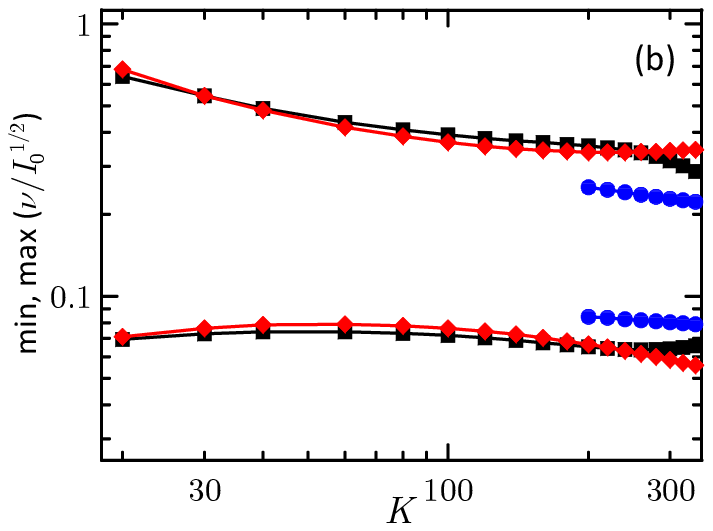}
\;
\includegraphics[width=0.32\textwidth]%
 {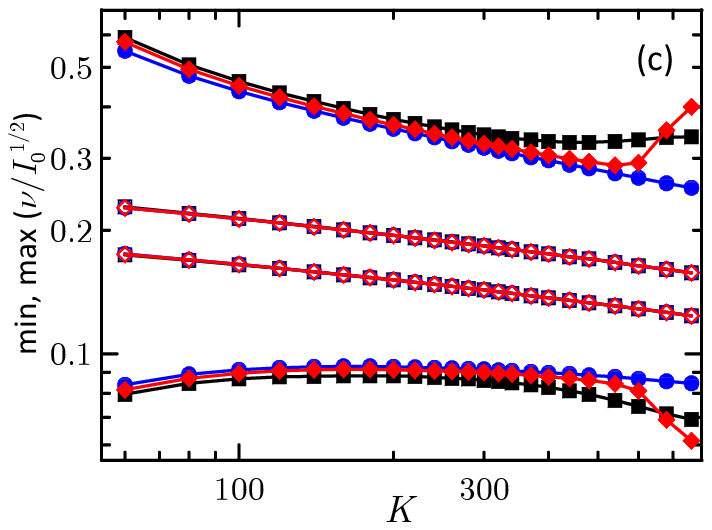}
}
\caption{Response of the network to the periodic modulation of excitatory current $\eta=0.4\cos{2\tau}$ is plotted with filled symbols for $i_0/g_0^2=0.2$~(a), $0.4$~(b), $0.8$~(c); black squares: the ``exact'' solution, red diamonds: 2CC reduction, blue circles: OA Ansatz. In panel~(b), the OA solution numerically explodes for $K<200$. In panel~(c), open symbols present the response for $\eta=0.4\cos{3\tau}$.
The values of $i_0$ and $K$ for the plots are shown with magenta lines in Figure~\ref{fig1}a.
}
  \label{fig3}
\end{figure}
\begin{figure}[!t]
\center{
\includegraphics[width=0.32\textwidth]%
 {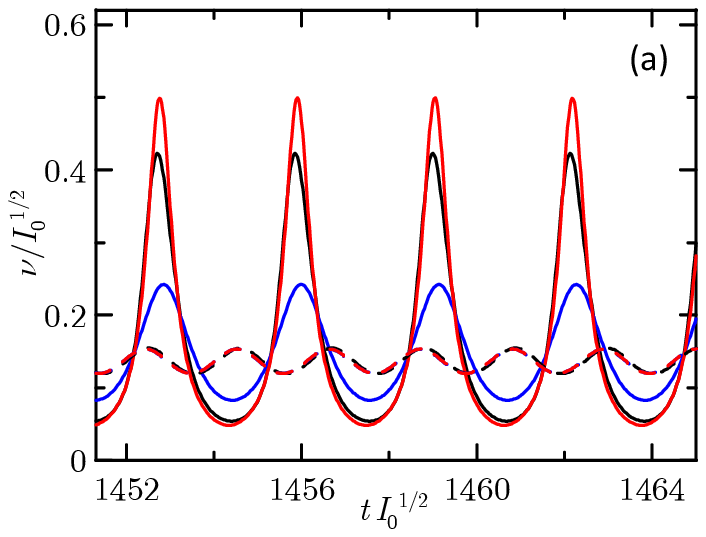}
\;
\includegraphics[width=0.32\textwidth]%
 {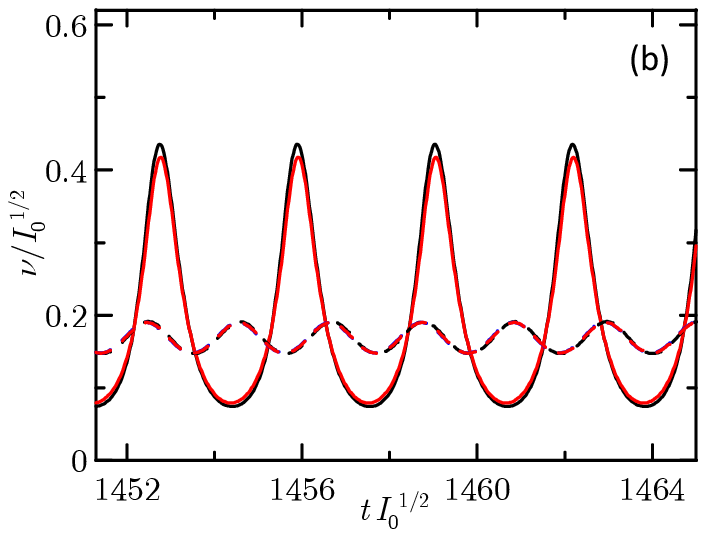}
\;
\includegraphics[width=0.32\textwidth]%
 {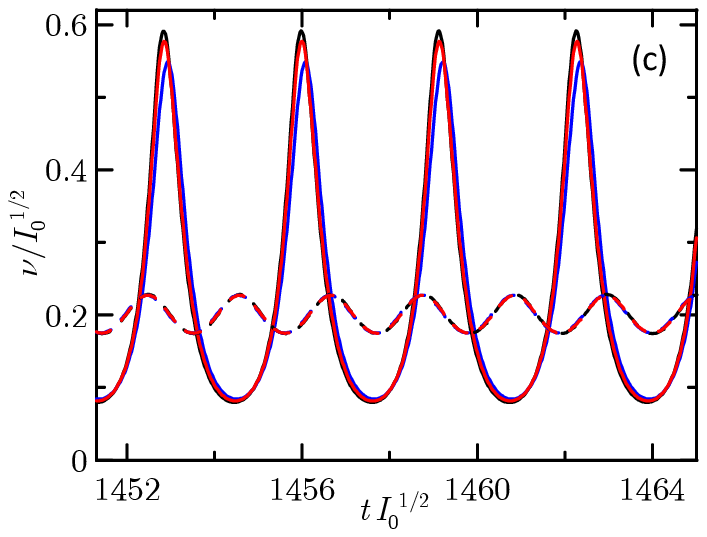}
}
\caption{Population firing rate versus time for $i_0$ and color coding as in Figure~\ref{fig3}, in- degree $K=60$, $\eta=0.4\cos{2\tau}$ (solid lines) and $0.4\cos{3\tau}$ (dashed lines).
}
  \label{fig4}
\end{figure}
\begin{figure}[!t]
\center{
\includegraphics[width=0.32\textwidth]%
 {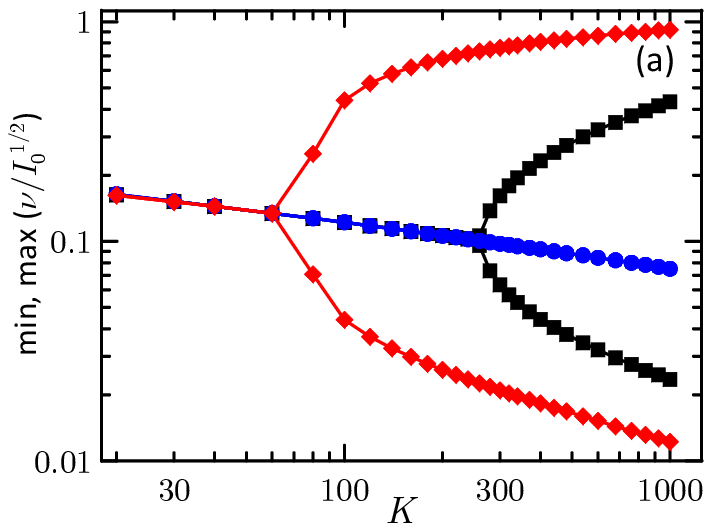}
\;
\includegraphics[width=0.32\textwidth]%
 {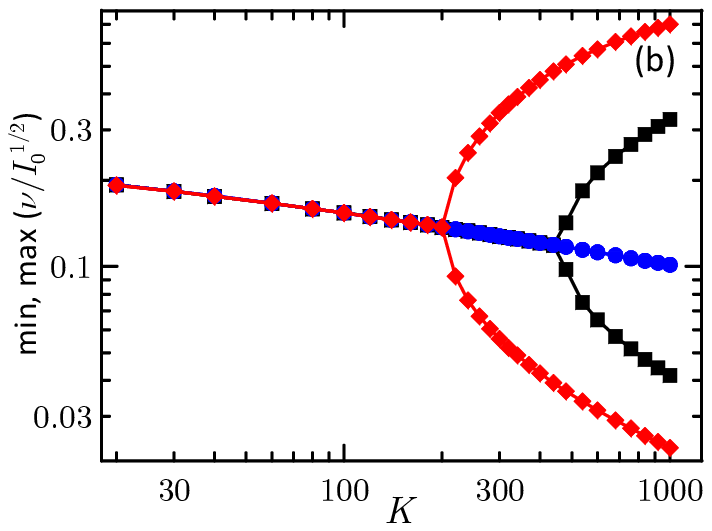}
\;
\includegraphics[width=0.32\textwidth]%
 {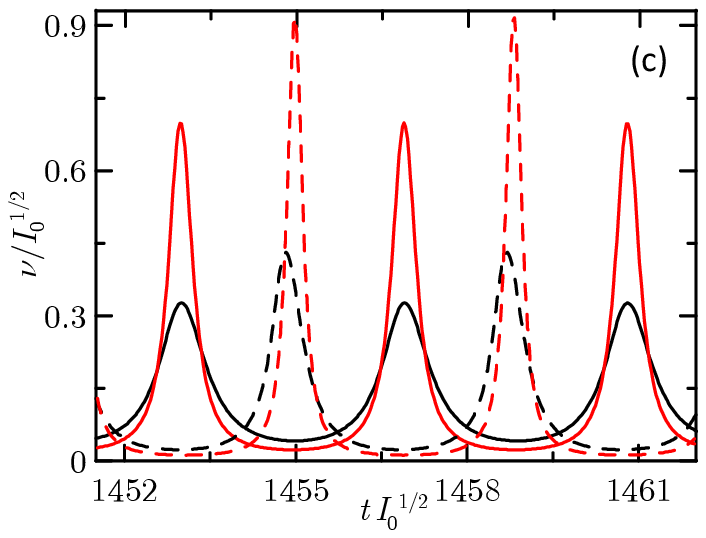}
}
\caption{For the driving-free system, the firing rate is plotted versus $K$ at $i_0/g_0^2=0.2$ (a) and $0.4$ (b) for the ``exact'' solution (black squares), 2CC reduction (red diamonds), and OA Ansatz (blue circles). In panel (c), the firing rate oscillations are plotted for $K=1000$, $i_0/g_0^2=0.2$ (dashed lines) and $0.4$ (solid); black: the ``exact'' solution, red: 2CC model.
}
  \label{fig5}
\end{figure}

In Figures~\ref{fig3} and \ref{fig4}, one can see that the 2CC model reduction decently captures the ``exact'' dynamic response to strong modulation $I(\tau)=I_0[1+0.4\cos\omega_\tau\tau]$ of the system for $i_0=0.2g_0^2$ and becomes much more accurate for higher excitatory currents. The OA Ansatz becomes applicable only for high currents $i_0$. Notice, $\omega_\tau=2$ is close to the resonant frequency of self-excited oscillations above the Hopf bifurcation threshold~\cite{Goldobin-etal-2024,Goldobin-etal-2025}, and the system dynamics is most sensitive here, making the simulations most demanding to the model accuracy. Away from the resonant frequency, e.g., for $\omega_\tau=3$ (see Figures~\ref{fig3}c and \ref{fig4}), the low dimensional models exhibit much higher accuracy and trajectories are indistinguishable from the ``exact'' solution.

In Figure~\ref{fig5}, for the driving-free system, the self-excitation of collective oscillations via a supercritical Hopf bifurcation can be witnessed from the dependence of the firing rate on $K$. In Figure~\ref{fig1}a, with the gray dash-dotted curve presenting the Hopf bifurcation threshold for the 2CC model reduction, one can see that the oscillatory instability threshold of this approximation is lowered in the parameter domain where the periodically driven oscillations are adequately reproduced by the 2CC model (i.e., for $i_0/g_0^2\gtrsim0.2$).

\section{Diffusion approximation}
\label{sec7}
2CC model reduction exhibits decent accuracy for time-independent regimes and dynamic response far beyond the domain of applicability of the diffusion approximation (DA) (see Figures~\ref{fig1}a). For the sake of completeness, in this Section we provide a 2CC model reduction for DA and derive the theoretical bounds for the applicability of this approximation (not just observation that it fails for important macroscopic regimes of the network~\cite{Goldobin-etal-2024,Goldobin-etal-2025}).

Consider the continuity equation (\ref{eqSN201}) rewritten with the diffusion approximation~\cite{Goldobin-etal-2024,Goldobin-etal-2025}:
\begin{equation}
\label{DA23:01}
{\partial_t P(V,t)} + {\partial_V}[(I(t) + V^2)P(V,t)] =
K\nu(t)\left[a\partial_{V}P(V,t) + \frac{a^2}{2}\partial^2_{V} P(V,t)\right] \; .
\end{equation}
Here one approximately represents the shot-noise term with the mean drift and continuous diffusion parts; technically, its Taylor expansion with respect to $a$ is truncated after the two leading terms.
For the probability density $w(\psi,t)$ of the genuine phase $\psi=2\arctan(V/\sqrt{I_0})$, continuity equation (\ref{DA23:01}) yields a modified version of (\ref{eqGP01})\;:
\begin{align}
\frac{\partial w(\psi,t)}{\partial t}
=-\frac{\partial}{\partial\psi}\left\{\sqrt{I_0}\big[2 +\eta(t)(1+\cos\psi)\big]w(\psi,t)\right\}
+K\nu(t)\left[\frac{a}{\sqrt{I_0}}\mathcal{Q}w(\psi,t) + \frac{a^2}{2I_0}\mathcal{Q}^2w(\psi,t)\right] \; ,
\label{DA23:02}
\end{align}
where the operator
\[
\mathcal{Q}(\dots)\equiv\frac{\partial}{\partial\psi}\Big[(1+\cos\psi)(\dots)\Big] \, .
\]
Hence, for the Kuramoto--Daido order parameters $z_n$, one finds a modified version of equation system (\ref{eqGP03eta}):
\begin{align}
&\dot{z}_n=in\sqrt{I_0}\left\{\big[2+\eta(t)\big]z_n
 +\eta(t)\frac{z_{n-1}+z_{n+1}}{2}\right\} +K\nu(t)\sum_{m=0}^{+\infty} I_{nm}^{[\mathrm{DA}]}z_m\;,
\quad n=1,2,3,\dots,
\label{DA23:03}
\end{align}
where $z_0=1$; the ``truncated'' matrix
$$
I_{nm}^{[\mathrm{DA}]}\equiv\alpha Q_{nm} +\frac{\alpha^2}{2}(\mathbf{Q}^2)_{nm}\,,
\qquad
\mathbf{Q}=\left(
\begin{array}{cccccc}
0 &0 &0 &0 &0 &\dots\\
-\frac{i}{2} & -i & -\frac{i}{2} & 0 &0 &\dots\\
0 & -i & -2i & -i & 0 &\dots\\
0& 0& -\frac{3i}{2} & -3i & -\frac{3i}{2} &\dots\\
&\dots&&
\end{array}
\right)\,,
\quad n=0,1,2,...,\;m=0,1,2,...\,.
$$
By substituting the matrix coefficients, one can recast Eq.~(\ref{DA23:03}) as follows:
\begin{align}
\dot{z}_n&
=in\sqrt{i_0}K^\frac14\left[2z_n +A(t)\left(z_n+ \frac{z_{n-1}+z_{n+1}}{2}\right)\right]
\nonumber\\
&
\quad{}
-\frac{n\nu(t)g_0^2}{2i_0\sqrt{K}}\left[\frac{n-1}{4}z_{n-2}+\left(n-\frac12\right)z_{n-1} +\frac{3n}{2}z_n +\left(n+\frac12\right)z_{n+1}+\frac{n+1}{4}z_{n+2}\right]\, ,
\label{DA23:04eta}
\end{align}
where $A(t)=\eta(t)-g_0\nu(t)/i_0$\,.
Similarly to~\cite{Volo-etal-2022}, one can write down the 2CC reduction for Eq.~(\ref{DA23:04eta}):
\begin{align}
\dot\varkappa_1&=i\sqrt{i_0}K^\frac{1}{4}\left\{2\varkappa_1 +A(t)\big[(1+\varkappa_1)^2+\varkappa_2\big]\right\}
-\frac{g_0^2\nu(t)}{4i_0\sqrt{K}} \left[(1+\varkappa_1)^3+3\varkappa_2(1+\varkappa_1)\right],
\label{DA23:05}
\\
\dot\varkappa_2&=i\sqrt{i_0}K^\frac{1}{4}\big[4\varkappa_2 +2A(t)(1+\varkappa_1)\varkappa_2\big]
 -\frac{g_0^2\nu(t)}{4i_0\sqrt{K}} \left[(1+\varkappa_1)^4+12\varkappa_2(1+\varkappa_1)^2\right].
\label{DA23:06}
\end{align}

\subsection{Theoretical boundary for applicability of the diffusion approximation}
\label{sec71}
Since the diffusion approximation is related to a truncated Taylor expansion of the noise term of Eq.~(\ref{eqGP01}) with respect to $a$, this approximation is valid as long as $a$ is small compared to the reference width $\sigma_V$ of the probability density distribution. Assuming the diffusion approximation is applicable and employing the results of~\cite{Volo-etal-2022}, we can estimate $\sigma_V$ for time-independent macroscopic states of Eq.~(\ref{DA23:01}). Indeed, Eq.~(\ref{DA23:01}) effectively describes the diffusion with coefficient $D=\nu g_0^2/2$ of an overdamped particle in the effective potential $U_\mathrm{eff}(V)=-A_gV-V^3/3$, $A_g\equiv\sqrt{K}(i_0-g_0\nu)$. The width $\sigma_V$ of the distribution in the local minimum of this potential can be estimated $\sigma_V^2=D/[2\sqrt{-A_g}]=g_0^2\nu/[4\sqrt{-A_g}]$. The quantitative measure for the approximation accuracy $a/\sigma_V$ should be small:
\[
\frac{a}{\sigma_V}\approx\frac{2(-A_g)^{1/4}}{\sqrt{K\nu}} \ll1\,.
\]
For small $(i_0/g_0^2)$ and time-independent states, asymptotic behavior of $A_g$ and $\nu$ was derived in~\cite{Volo-etal-2022}: $i_0/g_0^2\approx\nu/g_0-[(\nu/g_0)\ln(g_0/\nu)/2]^{2/3}/\sqrt{K}$. Whence one can find $A_g$ and estimate
\begin{equation}
\frac{a}{\sigma_V}\approx\frac{2[0.5\ln(g_0^2/i_0)]^{1/6}}{(i_0/g_0^2)^{1/3}\sqrt{K}}
=\sqrt{\frac{K_\mathrm{D}}{K}}\,.
\label{eqDA:small}
\end{equation}
This estimate gives an optimistic boundary of the applicability of the diffusive approximation, as the applicability can be also violated for time-dependent macroscopic regimes. Considering in- degree $K$, the approximation is accurate for $K$ 1--2 orders of magnitude larger than $K_\mathrm{D}=4[0.5\ln(g_0^2/i_0)]^{1/3}/(i_0/g_0^2)^{2/3}$ [notice the square root of $K$ in (\ref{eqDA:small}), which should be small].

\section{Conclusion}
\label{sec:concl}
The macroscopic dynamics of sparse balanced networks of neurons can be well represented with an effective synaptic shot noise driving a neuron~\cite{Volo-etal-2022,Goldobin-etal-2024,Goldobin-etal-2025} (see~\cite{Goldobin-etal-2025} for a concise review on biological relevance of the considered range of model parameters).
Most intriguing collective dynamics are observed in the parameter domain where one cannot adopt the diffusion approximation which is conventionally used for shot noise in mathematical neuroscience and physics of condensed matter.
The broad class of ``next generation neural mass models'' developed for no-noise/Cauchy noise and relatively recently generalized to the case of white Gaussian noise is inapplicable for a shot noise and demands an upgrade/adaptation for this case.
In this paper we have derived a low dimensional neural mass model for the case where the effect on intrinsic fluctuations is represented by an effective shot noise but essentially cannot be reduced to the diffusive approximation.

The model reduction is based on the circular cumulant formalism and, as a first step, requires the rewriting of the continuity equation in terms of the genuine phase. Further, we adopt a two circular cumulant truncation~(\ref{eqGPeta1})--(\ref{eqGPeta2}) for an infinite chain of circular moment equations~(\ref{eqGP03eta}). For completeness, we examined both the 2CC reduction and its further downgrade to the Ott--Antonsen Ansatz~\cite{Ott-Antonsen-2008,Ott-Antonsen-2009} by setting $\varkappa_2=0$.

For time-independent solutions the 2CC reduction is accurate for as low excitatory currents as $i_0=0.01g_0^2$, which extends by two orders of magnitude farther than the applicability domain of the diffusion approximation (Figure~\ref{fig1}a). For self-excited noise-induced oscillations the 2CC reduction is less accurate since it underestimates the diffusive suppression of collective modes and gives a lowered value of the Hopf bifurcation threshold as compared to the ``exact'' solution; for heterogeneous populations the threshold is more accurate~\cite{Goldobin-Volo-Torcini-2021,Volo-etal-2022}, but this is beyond the scope of our paper. Further, we analysed the accuracy of 2CC simulations for dynamic regimes (Figures~\ref{fig3} and \ref{fig4}). Even for a strong modulation of synaptic current with a resonant frequency, where the system is most sensitive to the problem of the overestimation of the noise-induced self-excitation of oscillations, the 2CC model captures the dynamic response for $i_0=0.2g_0^2$ and its error rapidly decreases as $i_0/g_0^2$ increases. For off-resonance frequencies, the dynamic response error is much smaller. Summarizing, the 2CC model might be expected to be applicable for theoretical studies of self-organized global oscillations in sparse networks where both excitatory and inhibitory synaptic links are present and collective oscillations emerge due to the interplay of excitation and inhibition~\cite{Vreeswijk-Sompolinsky-1996,Kadmon-Sompolinsky-2015,Atay-etal-2011,Segneri-Bi-Olmi-Torcini-2020,Bi-Volo-Torcini-2021,Kirillov-Zlobin-Klinshov-2023,Kirillov-Smelov-Klinshov-2024}.

An accurate agreement between the dynamics of the 2CC reduction and the complete mean-field model indicates and explains the low embedding dimensionality of attracting macroscopic regimes of the system both under constant and periodically modulated conditions.
Calculations with the derived 2CC reduction also do not require computation of coefficients $I_{nm}$. This is beneficial for instrumental applications since an accurate computation of $I_{nm}$ for high $n$ and $m$ requires an elevated accuracy of the floating point operations~\cite{Goldobin-etal-2025}.

The Ott--Antonsen reduction of the 2CC model has been found to be accurate only slightly beyond the applicability domain of the diffusion approximation. Moreover, it is completely unable to represent the effect on noise-induced self-excitation of collective oscillations (the Lyapunov exponent $\lambda$ in Figure~\ref{fig1}c is always negative). Comparing the 2CC reduction for DA~(\ref{DA23:05})--(\ref{DA23:06}) with the 2CC reduction for shot noise~(\ref{eqGPeta1})--(\ref{eqGPeta2}), one can see that the diffusion terms are different not only for the 2CC models but also for their OA reductions obtained by setting $\varkappa_2=0$.

\vspace{-2mm}
\paragraph{Acknowledgements} The authors are thankful to A.\ Torcini for discussions.
The work was carried out as part of a major scientific project (Agreement No.\ 075-15-2024-535 by 23 April 2024).

\vspace{-2mm}
\paragraph{Author contributions}
All authors equally contributed to the study conception and design, theoretical derivations, coding, numerical simulations and data analysis. The first draft of the manuscript was written by M.V.~Ageeva, the draft was finalized by D.S.~Goldobin. All authors read and approved the final manuscript.

\vspace{-2mm}
\paragraph{Data availability} All data generated and analyzed during this study are included in this paper.


\vspace{-2mm}
\paragraph{Conflict of interest} The authors declare that they have no conflict of interest.

\end{document}